\documentclass[%
mathleft,%
]{an}
\usepackage{graphicx}
\usepackage{times}
\begin{document}

\Pagespan{1}{}
\Yearpublication{2011}%
\Yearsubmission{2011}%
\Month{1}%
\Volume{999}%
\Issue{92}%
 \DOI{This.is/not.aDOI}%

\title{Starspots and spin-orbit alignment for Kepler cool host stars}

\author{R. Sanchis-Ojeda\inst{1}\fnmsep\thanks{Corresponding author:
  \email{rsanchis86@gmail.com}}
\and  J.\,N. Winn\inst{1}
\and  D.\,C. Fabrycky\inst{2}
}
\titlerunning{Starspots and spin-orbit alignment for Kepler cool host stars}
\authorrunning{R. Sanchis-Ojeda, J.\,N. Winn \& D.\,C. Fabrycky}
\institute{
Department of Physics, Massachusetts Institute of Technology, Cambridge, Massachusetts 02139, USA.
\and 
Department of Astronomy and Astrophysics, University of California, Santa Cruz, Santa Cruz, California 95064, USA.}

\received{XXXX}
\accepted{XXXX}
\publonline{XXXX}


\abstract{The angle between the spin axis of the host star and the orbit of its planets (i.e., the stellar obliquity) is precious information about the formation and evolution of exoplanetary systems. Measurements of the Rossiter-McLaughlin effect revealed that many stars that host a hot-Jupiter have high obliquities, suggesting that hot-Jupiter formation involves excitation of orbital inclinations. In this contribution we show how the passage of the planet over starspots can be used to measure the obliquity of exoplanetary systems. This technique is used to obtain - for the first time - the obliquity of a system with several planets that lie in a disk, Kepler-30, with the result that the star has an obliquity smaller than 10 degrees. The implications for the formation of exoplanetary systems, in particular the hot-Jupiter population, are also discussed. }

\maketitle

\section{Introduction}

A commonly used technique to measure the obliquity of exoplanet systems is the  Rossiter-Mclaughlin (RM) effect (Queloz et al. 2000, Winn et al. 2005), a spectroscopic phenomenon that is observed during transits. Since the planet blocks a certain part of the rotating stellar surface, an additional Doppler shift is observed which depends on the sky-projected obliquity of the system. Thanks to this effect, the obliquity of many hot-Jupiters systems has been found to be rather large (Winn et al. 2010a; Triaud et al. 2010; Albrecht et al. 2012), a somehow surprising discovery since our own Sun is known to have a low obliquity. When a solar system is born through the collapse of a cloud of gas, the star and the protoplanetary disk are expected to rotate in the same direction. So the question is, how did the hot-Jupiters get misaligned?

There are two possible reasons to explain why the hot-Jupiters have been found to be misaligned. One is that the spin axis of the star can get tilted with respect to the original position of the protoplanetary disk, either through magnetic interactions (Lai, Foucart \& Lin 2011), chaotic accretion (Bate, Lodato \& Pringle 2010) or torques from neighbor stars. In this scenario the hot-Jupiters migrate inwards on the disk to get to their close-in orbits, and the high obliquities observed are a consequence of the star-disk misalignment. The other possible explanation is that dynamical interactions, such as planet-planet scattering (Rasio \& Ford 1996) or Kozai cycles (Fabrycky \& Tremaine 2007), can tilt the orbits of the planets. During Kozai cycles orbital eccentricities can reach very large values, leading to small periastron distances during phases of high eccentricities. Tidal interactions then circularize and shrink the orbit, explaining the close-in orbits of the hot-Jupiters. In this case, the high obliquities observed are a consequence of the history of dynamical interactions in the system. 

This raises a new question, how are we going to distinguish between these two theories? The answer necessarily implies studying the obliquities outside the group of the hot-Jupiters. We propose to study the multiple transiting systems for several reasons. These systems are more likely to be coplanar, simply because they are all transiting the same star (Lissauer et al. 2011). If a system has several coplanar planets, it probably had a relatively quiet dynamical history. The second explanation for the high obliquities of hot-Jupiter systems needs these dynamical interactions to tilt the orbits, so if this scenario is correct it predicts that all the coplanar transiting systems should be aligned, similar to our own Solar system. However, the first explanation, which invokes star-disk misalignments, can be in principle extended to all types of solar systems. Even in this coplanar configuration, the star can still get misaligned respect to the disk, so misalignment should be observed with a similar frequency in these systems. 

Now we have a clear way to distinguish between both theories, but the observational challenge is great. Almost all multiple transiting planets found to date have been discovered by the \emph{Kepler} mission (Batalha et al. 2012), but  follow up of Kepler targets tends to be challenging due to the faintness of the host stars (typical \emph{V} magnitudes of 13-16). In addition, the RM effect scales with the projected rotation speed of the host star, but many of the systems have late type stars which are slow rotators. For both of these reasons, RM measurements for multiple-planet systems have been difficult, leading us to seek an alternative means of measuring stellar obliquities.

\section{Using starspots to obtain the obliquity}

\begin{figure}
\center

\includegraphics[keepaspectratio,height=105mm]{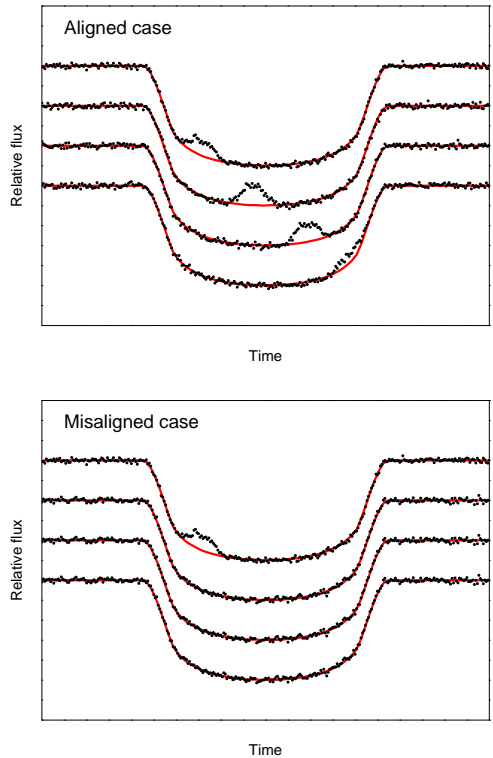}
\caption{A sequence of four simulated consecutive transits shows the effect of a single spot in an aligned system (upper panel) and a misaligned system (lower panel). The rotation of the star is ten times slower than the orbit of the planet. When the system is aligned, the spot can be followed in consecutive transits. When the system is misaligned, spots appear in single events and they are not seen in the following transits. }
\label{label1}
\end{figure}

In the past few years a new technique has appeared with the goal of expanding the obliquity measurements in parameter space (towards longer orbital periods and smaller planets). The technique uses the passage of the planet over starspots to obtain information about the obliquity. It is complementary to the RM effect because it works well for slow rotators and can be used for faint stars, so long as the starspots are prominent enough. The essence of the technique is the following: if a system has a low obliquity, an occulted spot should be occulted in consecutive transits (see Fig. \ref{label1}). Imagine that we have a star with a transiting planet. The part of the surface of the star transited by the planet is called the transit chord. Now if a dark starspot is on the transit chord when the planet is transiting, the planet will pass in front of it, blocking less light than expected for that certain part of transit. This temporary brightening will be seen as a bump on the light curve, something that has been observed many times in high photometric precision transits (Rabus et al. 2009). In the case of a perfectly aligned system, with a slowly rotating host star, when the transiting planet completes an entire orbit and comes back to transit, the spot will still be on the transit chord. This is true because the trajectory of the spot is parallel to the trajectory of the planet. But the rotation of the star will make the spot advance on the transit chord. The result is that the bump appears again during the next transit, but its position also advances accordingly. This will happen for a few more transits until the spot is taken to the hidden side of the star. 

In the case of a misaligned system, the pattern is very different as the 
star spots do not travel parallel to the transit chord. If a spot is 
covered during one transit by the planet then it will have moved away 
from the transit cord by the time of the next transit -- it will be 
missed completely.

The idea of tracking spots during transits was first used to estimate stellar rotation periods (Silva-Valio 2008, Dittmann et al. 2009). Later, the technique was first applied to measure obliquities on the hot-Jupiter system WASP-4, which was shown to be aligned thanks to ground based photometry when two spots were seen twice in different transits (Sanchis-Ojeda et al. 2011). Similarly, the hot-Jupiters CoRoT-2b (Nutzman, Fabrycky \& Fortney 2011) and Kepler-17b (D{\'e}sert et al. 2011) were also shown to be aligned. These latter analyses were based on data gathered from space telescopes, and they show how important continuous coverage can be. Nutzman, Fabrycky \& Fortney (2011) showed how by modelling the stellar flux variations with several spots, one can find the spot responsible for the spot-crossing events. For that particular spot, the longitudes inferred from the stellar flux variations can be easily transformed into a spot anomaly phase along the transit chord only in the case of low obliquity. 

\section{A high obliquity for HAT-P-11 and the butterfly diagram}

The method was then applied to a misaligned case, the hot-Neptune system HAT-P-11.  The system was discovered on the Kepler field before the beginning of the Kepler mission (Bakos et al. 2010). The RM effect detection shows that system is highly inclined (Winn et al. 2010b, Hirano et al. 2011). The almost continuous coverage of the Kepler data showed clear stellar flux variations, and confirmed a rotation period of about 30 days (Sanchis-Ojeda \& Winn 2011). Spot-crossing events are frequent during the transits of HAT-P-11b, but the recurrences are not observed. Spots are not followed in consecutive transits, the system must be misaligned. A quick look at the folded transit light curve shows that the spot-anomalies only appear at two very specific phases of the transit (Sanchis-Ojeda \& Winn 2011; Deming et al. 2011). We interpret this as a sign that the star has active latitudes, where spots appear. This behaviour is characteristic of our own Sun, where the active latitudes are symmetric respect to the equator. Since the system is misaligned, the planet is crossing a wide range of latitudes, but can only encounter spots at those latitudes that are active. Using this argument, we showed that the system must have an obliquity of about $100^{\circ}$ (Sanchis-Ojeda \& Winn 2011), in agreement with the RM effect. For this system, and misaligned systems in general, we can in principle extract the latitude of individual spots once the obliquity of the system is known. For each system we can probe a certain range of latitudes that depends on the geometry of the system. Changes in the latitudes of sunspots correlate with the magnetic cycle of the Sun, in what is called the butterfly diagram, so we might be able to construct the butterfly diagram for HAT-P-11 and other misaligned system if we observe temporal structure on the latitudes of the occulted spots. 

\section{The obliquity of Kepler-30, a coplanar system with three transiting planets}

Now that we have checked that the technique works for gas giants really close to their host stars, it is time to apply it to systems with several transiting planets a little more distant from the host star, which orbits are tracing the plane of the disk out of which they formed.

In order to apply this method, we need the individual transits to be significant enough for the detection of the spot-anomalies and we also need the host star to be active. Only a few candidates from the Kepler multiple transiting systems fit our requirements. One of the best, Kepler-30, was the one chosen for this study. Kepler-30 is a system with three planets transiting a Sun-like star (Fabrycky et al. 2012).  Planet b is Neptune-sized and has an orbital period of about 29.3 days, and planet c and d are Jupiter-sized with orbital periods of 60.3 and 143.3 days. The planets are much further away from the host star than Hot-Jupiters, and tidal interactions are negligible. The two outer planets are large enough to show the anomalies. The first thing we did was to look for recurrences. The problem is that long period planets have very few transits during the lifetime of a spot, even in the case of large spots that can last for months. In this particular case, the rotation period of the star is $16.0 \pm 0.4$ days, obtained from a periodogram analysis.  The star completes $3.77 \pm 0.09$ rotations in between two consecutive transits of Kepler-30c. 

We searched carefully the 12 transits of Kepler-30c available and found a recurrence. Figure 2 of Sanchis-Ojeda et al. (2012) shows two spot-crossing events in two consecutive transits of Kepler-30c. During the first transit, the planet hits a rather large spot on the latter half of the transit. With one full orbital period of planet Kepler-30c, the star has rotated 3.77 times, which is equivalent to a backwards rotation of $80^{\circ}$. If the system has a low obliquity, when the planet transits again it should hit the spot on the left side of the star, that is, on the first half of the transit. And that is what we observe. In addition to this, Kepler-30d had a transit 16 days after the second transit of Kepler-30c, and it seems that this other planet transited in front of the same spot with the same phase. In order to believe this interpretation, we need to make sure that we are talking about the same spot on the three cases.

The benefit of using this technique for Kepler or CoRoT targets is that the continuous monitoring of the stellar flux provides extra information about the spots. Since the spot is quite large, it must contribute strongly to the variations seen in the disk-integrated light outside of transits. If the system has a low obliquity, as we are claiming, there should be a deep local flux minimum when the spot lies on the longitude that crosses through the center of transit chord. Observing the spot on the left (right) side of the transit means that the spot is coming into (out of) view, and that the transit will happen before (after) the corresponding deep local flux minimum. 

And that is exactly what is observed when one looks at the flux of the star during the 100 days of relevant Kepler observations (Sanchis-Ojeda et al. 2012). All three transits have the right relationship, with transits to the right of a flux minima that show the large anomaly on the right side and vice versa.  Furthermore, all the relevant flux minima belong to a series of deep flux minima that appear periodically every 16 days, the rotation period. This makes it very easy to explain all the observations with a large spot, with a rotation period of 16 days, with a lifetime larger than 100 days, which has a trajectory parallel to the trajectory of the planet, so that the spot is hit at the right phase in each of the three transits. This is further evidence that the system has a low obliquity.

\begin{figure}
\center
\includegraphics[keepaspectratio,height=125mm]{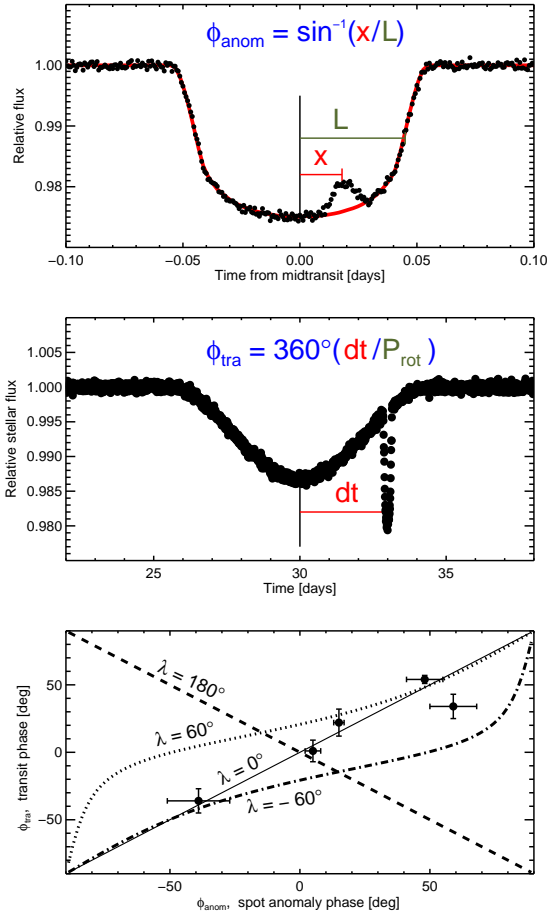}
\caption{Upper panel: definition of the anomaly phase, which can be read directly from the transit light curve. Middle panel: definition of the transit phase, which can be read from the stellar flux variations. Lower panel: their relationship for different values of the sky-projected obliquity represented with different types of lines. In Kepler-30 the phases are equal, and this can only be consistently explained if the obliquity is low (thick line). }
\label{label2}
\end{figure}

We would like to confirm this with additional information from the other transits. Since we identified no more recurrences, we have developed a new technique to use the information from spot-crossing events of spots that are only observed once. Using what Nutzman et al. (2011) applied to study CoRoT-2b, we define two different angles or phases, one that can be measured from the transit, and one that can be measured from the stellar flux (see Fig. \ref{label2}). The first phase is the anomaly phase, measured along the transit chord, which goes from $-90^{\circ}$ on the ingress to $90^{\circ}$ at egress.

The other phase, called transit phase, is basically the stellar longitude of the spot at the time of transit. This angle is linear with time, and is zero when the flux drop due to the particular spot is maximum, generating the characteristic local flux minima. Therefore, the transit phase is the difference in time between the transit and the correspondent flux minimum, normalized by the rotation period. 

Now we take the step further to assume that both phases measured belong to the same spot. It is crucial to select the spot-crossing anomalies we are working with, so that they are caused by very large spots. This ensures that these spots also cause large stellar flux variations, so we can measure both phases securely. We also need to be careful with Kepler systematics, and avoid transits that are close to large data gaps. These criteria leave only 5 individual useful events, two of which are nothing but the recurrence of the same spot already analized. For all of them, we measure the anomaly phases and transit phases carefully, using MCMC algorithms to properly estimate the associated errors. 

We now plot the results (see Fig. \ref{label2}), with the anomaly phase on the x axis and the transit phase relative to the flux minimum on the y axis. The black dots represent the observations with the associated error bars. The lines represent the relationship between both phases for different sky-projected obliquities, with a fixed inclination of the star of $90^{\circ}$. The black line, the zero obliquity one, gives the best fit for all the points, showing further evidence that the system is well aligned. In addition to this result, in the Supplentary information of our paper (Sanchis-Ojeda et al. 2012) we show how to exhaustively deal with degeneracy problems associated with the multiplicity of spots observed on the stellar flux variations. We also show that the three planets are coplanar, since mutually inclined orbits would make the system rapidly precess and the resulting drifts in duration have not been detected. Finally, we also obtain the masses and radius of the planets. 

\section{Conclusions}

In this contribution we have shown that transits over starspots can be used to measure the obliquities of exoplanet systems and to obtain the butterfly diagram of the host star. The low obliquity of Kepler-30 is the first measurement of the obliquity of a system outside the hot gas giants population. This low obliquity, together with the low obliquity of our own Solar System, suggests that the high obliquities are confined to hot-Jupiter systems, which if confirmed would leave few-body dynamics as the origin of the hot-Jupiter population. Additional stars with coplanar multiplanet systems should be observed, with the spots technique or others, and are predicted to have low obliquities in order for our conclusions to hold.

\acknowledgements

We would like to thank Simon Albrecht for discussing this document. We would also like to thank all the members of the Kepler team who made the mission possible, especially to the coauthors of our Kepler-30 paper. Finally, we would like to thank the organizers of Cool Stars 17 for a great conference.

%

\begin{thebibliography}{}

\bibitem[Albrecht et al.(2012)]{2012ApJ...757...18A} Albrecht, S., Winn, 
J.~N., Johnson, J.~A., et al.\ 2012, ApJ, 757, 18 

\bibitem[Bakos et al.(2010)]{2010ApJ...710.1724B} Bakos, G.~{\'A}., Torres, 
G., P{\'a}l, A., et al.\ 2010, ApJ, 710, 1724 
\bibitem[Batalha et al.(2012)]{2012arXiv1202.5852B} Batalha, N.~M., Rowe, 
J.~F., Bryson, S.~T., et al.\ 2012, arXiv:1202.5852 

\bibitem[Bate, Lodato \& Pringle (2010)]{2010MNRAS.401.1505B} Bate, M.~R., Lodato, G., 
\& Pringle, J.~E.\ 2010, MNRAS, 401, 1505 

\bibitem[Deming et al.(2011)]{2011ApJ...740...33D} Deming, D., Sada, P.~V., 
Jackson, B., et al.\ 2011, ApJ, 740, 33 
\bibitem[D{\'e}sert et al.(2011)]{2011ApJS..197...14D} D{\'e}sert, J.-M., 
Charbonneau, D., Demory, B.-O., et al.\ 2011, ApJS, 197, 14 
\bibitem[Dittmann et al.(2009)]{2009ApJ...701..756D} Dittmann, J.~A., 
Close, L.~M., Green, E.~M., \& Fenwick, M.\ 2009, ApJ, 701, 756 
\bibitem[Fabrycky \& Tremaine(2007)]{2007ApJ...669.1298F} Fabrycky, D., \& Tremaine, S.\ 2007, ApJ, 669, 1298 
\bibitem[Fabrycky et al.(2012)]{2012ApJ...750..114F} Fabrycky, D.~C., Ford, 
E.~B., Steffen, J.~H., et al.\ 2012, ApJ, 750, 114 

\bibitem[Hirano et al.(2011)]{2011PASJ...63S.531H} Hirano, T., Narita, N., 
Shporer, A., et al.\ 2011, PASJ, 63, 531 
\bibitem[Lai, Foucart \& Lin (2011)]{2011MNRAS.412.2790L} Lai, D., Foucart, F., 
\& Lin, D.~N.~C.\ 2011, MNRAS, 412, 2790 

\bibitem[Lissauer et al.(2011)]{2011ApJS..197....8L} Lissauer, J.~J., 
Ragozzine, D., Fabrycky, D.~C., et al.\ 2011, ApJS, 197, 8 
\bibitem[Nutzman, Fabrycky \& Fortney (2011)]{2011ApJ...740L..10N} Nutzman, P.~A., 
Fabrycky, D.~C., \& Fortney, J.~J.\ 2011, ApJ, 740, L10 
\bibitem[Queloz et 
al.(2000)]{2000A&A...359L..13Q} Queloz, D., Eggenberger, A., Mayor, M., et al.\ 2000, A\&A, 359, L13 
\bibitem[Rabus et al.(2009)]{2009A&A...494..391R} Rabus, M., Alonso, R., Belmonte, J.~A., et al.\ 2009, A\&A, 494, 391 
\bibitem[Rasio \& Ford(1996)]{1996Sci...274..954R} Rasio, F.~A., \& Ford, E.~B.\ 1996, Science, 274, 954 


\bibitem[Sanchis-Ojeda et al.(2011)]{2011ApJ...733..127S} Sanchis-Ojeda, 
R., Winn, J.~N., Holman, M.~J., et al.\ 2011, ApJ, 733, 127  
\bibitem[Sanchis-Ojeda \& Winn(2011)]{2011ApJ...743...61S} Sanchis-Ojeda, R., \& Winn, J.~N.\ 2011, ApJ, 743, 61 
\bibitem[Sanchis-Ojeda et al.(2012)]{2012Natur.487..449S} Sanchis-Ojeda, 
R., Fabrycky, D.~C., Winn, J.~N., et al.\ 2012, Nature, 487, 449 
\bibitem[Silva-Valio(2008)]{2008ApJ...683L.179S} Silva-Valio, A.\ 2008, 
ApJ, 683, L179 
\bibitem[Triaud et al.(2010)]{2010A&A...524A..25T} Triaud, A.~H.~M.~J., Collier Cameron, A., Queloz, D., et al.\ 2010, A\&A, 524, A25 

\bibitem[Winn et al.(2005)]{2005ApJ...631.1215W} Winn, J.~N., Noyes, R.~W., 
Holman, M.~J., et al.\ 2005, ApJ, 631, 1215 
\bibitem[Winn et al.(2010a)]{2010ApJ...718L.145W} Winn, J.~N., Fabrycky, D., 
Albrecht, S., \& Johnson, J.~A.\ 2010, ApJ, 718, L145 
\bibitem[Winn et al.(2010b)]{2010ApJ...723L.223W} Winn, J.~N., Johnson, 
J.~A., Howard, A.~W., et al.\ 2010, ApJ, 723, L223
\end{thebibliography}
%

\end{document}